# Interface-induced spin-resolved type-II band alignment and enhanced magnetic anisotropy in $MSe_2/WTe_2$ (M = V, Cr, Mn, Fe and Co) van der Waals Heterobilayers

Paras Poswal,[1,†] Ranjit Pati,[2] and Neeraj Shukla[1,*]

[1]Department of Applied Physics and Materials Engineering, National Institute of Technology Patna, Bihar 800005, India

[2]Department of Physics, Michigan Technological University, Houghton, MI, 49931, United States

**ABSTRACT.**

Two-dimensional van der Waals heterobilayers provide an attractive platform for the development of next-generation spintronic devices. Here, first-principles calculations are performed to investigate the structural, electronic and magnetic properties of $MSe_2/WTe_2$ (M = V, Cr, Mn, Fe, and Co) van der Waals heterobilayers. The pristine $WSe_2/WTe_2$ heterobilayer in AA′-configuration is found to be energetically favorable and exhibits type-II band alignment with a band gap of 0.70 eV, and this provides an ideal platform for controlling carrier transport. Substituting W with 3d transition metal atoms, induces long-range magnetic ordering and reconstructs the spin-resolved electronic band structure. The formation of the heterointerface generates pronounced charge redistribution and an intrinsic built-in electric field, leading to interface-induced electronic reconstruction. $MnSe_2/WTe_2$ heterobilayer exhibits half-metallicity, whereas $FeSe_2/WTe_2$ heterobilayer simultaneously exhibits half-metallicity and spin-resolved type-II band alignment. Interfacial electronic reconstruction further produces a substantial perpendicular magnetic anisotropy, driving $MnSe_2$ from an in-plane easy axis with MAE value of 1.10 meV in the isolated monolayer to a robust out-of-plane easy axis with MAE value of 20.8 meV in the heterobilayer. Among all the structures, $CoSe_2/WTe_2$ heterobilayer exhibits maximum Curie temperature (273.87 K). The combined results establish that interface engineering makes $MSe_2/WTe_2$ heterobilayers as a promising candidates for next-generation low-dimensional spintronic applications.

## I. INTRODUCTION

The stabilization of long-range magnetic ordering at finite temperature in two-dimensional (2D) materials remains one of the key challenges. This fundamental limitation has been established by Mermin and Wagner in 1966. According to Mermin-Wagner theorem, in an infinitely extended 2D system, which possesses continuous spin-rotational symmetry, long-range magnetic ordering cannot be sustained at finite temperature due to thermal spin fluctuations [1]. The presence of magnetic anisotropy in the system breaks the continuous spin-rotational symmetry and opens a finite gap in the spin-wave spectrum, which suppresses long-wavelength or low-energy magnon excitation and stabilizes long-range magnetic order in 2D materials at finite temperature [2,3].

The experimental observation of intrinsic ferromagnetic ordering in monolayer $CrI_3$ with curie temperature ~45 K attracted considerable attention in scientific community [4]. This discovery triggered the exploration of ferromagnetism in two-dimensional materials such as, $Cr_2Ge_2Te_6$ [5], $Fe_3GeTe_2$ [6]etc. One of the fascinating properties of these 2D magnetic materials is layer-dependent magnetic ordering. $CrI_3$ exhibits a ferromagnetic ground state in odd-layered structures, whereas even-layered structures favor an antiferromagnetic ground state [4]. Deng *et al.* reported that the Curie temperature of $Fe_3GeTe_2$ decreases from 180 K in the bulk to 20 K in the monolayer limit [7]. The relatively low Curie temperatures and modest magnetic anisotropy energies of these systems continue to limit practical spintronic applications. The flexibility of forming heterostructures of 2D materials offers additional advantages to overcome these limitations through interfacial engineering. Variations in stacking order and relative twist angles between the layers can significantly modify the band alignment, magnetic interactions, and spin polarization [8,9].

Within the broad class of 2D materials, TMDs ($MX_2$: M= Transition metal, X=chalcogen) have emerged as essential building blocks due to their structural stability, high spin-orbit coupling (SOC), valley polarization and highly tunable band structure [10]. The layered structure of TMDs having weak interlayer vdW interaction and intralayer covalent bonding, allow facile layer exfoliation and heterostructure formation down to the monolayer limit [11]. Among all TMDs, tungsten-based TMDs attracts considerable attention due to their heavy atomic mass. In the $WX_2$ family, $WSe_2$ has been shown to exhibit strong SOC-induced band splitting, enabling spin-selective carrier transport and valleytronics functionalities without the need for an external magnetic field [12–14]. Beyond $WSe_2$, the incorporation of heavier chalcogen atom (Te) further enhances the electronic properties of tungsten dichalcogenides. $WTe_2$ exhibits a variety of fascinating quantum phenomena such as type-II Weyl semimetal behavior, nontrivial topological states and giant magnetoresistance, which arise from its strong SOC [15,16]. Forming a van der Waals heterostructure by combining Se-based and Te-based TMDs creates a unique interfacial environment that can substantially modify the electronic structure, enhance magnetic anisotropy, and induce spin splitting. Recent studies on TMD-based heterostructure have demonstrated that heterostructures of $WTe_2$ with other TMDs or magnetic 2D materials hold great potential for flexible electronics and spintronics [17,18]. In this work, we have considered $MSe_2$ and $WTe_2$ based heterobilayers to explore magnetic ordering. The chemical asymmetry between the Se and Te chalcogen layers inherently breaks the inversion and mirror symmetries, while substitution of 3*d* Transition Metal (TM) atoms at the W site breaks time-reversal symmetry and induces magnetism. This combination is expected to enhance magnetic anisotropy, stabilize long-range magnetic ordering in the two-dimensional limit, and establish these heterobilayers as promising candidates for spintronic applications.

*Contact author: neerajs@nitp.ac.in,

†Contact author: parasp.ph21.ph@nitp.ac.in

## II. COMPUTATIONAL METHODS

### A. Density Functional Theory

In the present work, Quantum ESPRESSO [19] open-source package has been employed to perform first-principles density functional theory (DFT) calculations. The Generalized Gradient Approximation (GGA) with Perdew-Burke-Ernzerhof (PBE) [20] functional has been considered to describe the exchange-correlation interaction. The electron-ion interaction has been described using scaler-relativistic Projector Augmented-Wave (PAW) pseudopotential [21,22]. The kinetic energy cutoff of 70-80 Ry for the wave functions and 700-800 Ry for charge density cutoff has been considered. The Brillouin zone has been sampled using $9 \times 9 \times 1$ Monkhorst-Pack [23] k-point mesh, for structural optimization, and dense $12 \times 12 \times 1$ Monkhorst-Pack k-point mesh for electronic structure calculations. All atomic positions and lattice parameters have been relaxed until the Hellmann-Feynman force on each atom is less than $10^{-4}$ Ry/Bohr with total energy convergence threshold of $10^{-8}$ Ry. A vacuum exceeding 30 Å has been added along the z-axis to eliminate the artificial interaction between the periodically repeated images of the structure. The van der Waals interactions have been considered by Grimme DFT+D3 correction [24]. In order to accurately consider the localized nature of the 3d electrons, the Hubbard correction (DFT+U) has been considered. The Hubbard U-value has been calculating using linear response method [25]. The magnetic ground state energy of magnetic configurations has been calculated by enabling spin-polarized calculation. A dense $49 \times 49 \times 1$ Monkhorst-Pack k-point mesh, with fully relativistic PAW pseudopotential enabling spin-orbit coupling (SOC) and noncollinear magnetism has been considered to evaluate Magnetic Anisotropy Energy (MAE).

### B. Energy Mapping method

To determine the magnetic ordering and quantify the magnetic exchange interaction strength in heterobilayers system, the magnetic exchange coupling parameter has been calculated utilizing the total-energy mapping method. The magnetic interactions between the nearest-neighbor magnetic atoms have been described within the Heisenberg Hamiltonian [26,27],

$$H = -\sum_{i,j} J_{ij} \boldsymbol{S_i}.\boldsymbol{S_j} \tag{1}$$

Where, $J_{ij}$ denotes the exchange coupling constant between magnetic atoms at lattice sites *i* and *j*, and $\boldsymbol{S_i}$ and $\boldsymbol{S_j}$ are the corresponding spin vectors. A 2×2×1 supercell has been considered to calculate the total energy of ferromagnetic (FM) and antiferromagnetic (AFM) configuration, as shown in Fig. S1. If we consider only the dominant nearest neighbor interaction (*J*), the total energy of the different magnetic configurations (FM, AFM) can be expressed as,

$$E_{FM} = E_0 - 12S^2J \tag{2}$$

$$E_{AFM} = E_0 + 8S^2J - 4S^2J \tag{3}$$

Utilizing Eqs. (2) to (3), J can be expressed as,

$$J = (E_{AFM} - E_{FM})/16S^2 \tag{4}$$

Where, $E_{FM}$ and $E_{AFM}$ denote the total energy for the FM and AFM configuration, with non-magnetic energy contribution $E_0$, respectively.

### C. Monte-Carlo Simulation

To investigate the finite-tempertaure magnetic behaviour and magnetic phase transition temperature of $MSe_2/WTe_2$ hetrobilayer, atomistic spin simulations has been performed using VAMPIRE simulation package [26]. This simulation package is based on classical Heisenberg model and utilizes the calculated DFT-driven exchange parameters. The simulation have been perfomed on 15 nm × 15 nm × 4.0 nm supercell, which contains approximately 1849 magnetic atoms to minimize the finite size effects. The Metropolis Monte Carlo algorithm [28] has been considered to find out the temperature dependent magnetic properties. For each temperature, the system has been first equilibrated for 10,000 Monte Carlo steps, followed by an average run of 5,000 Monte Carlo steps.

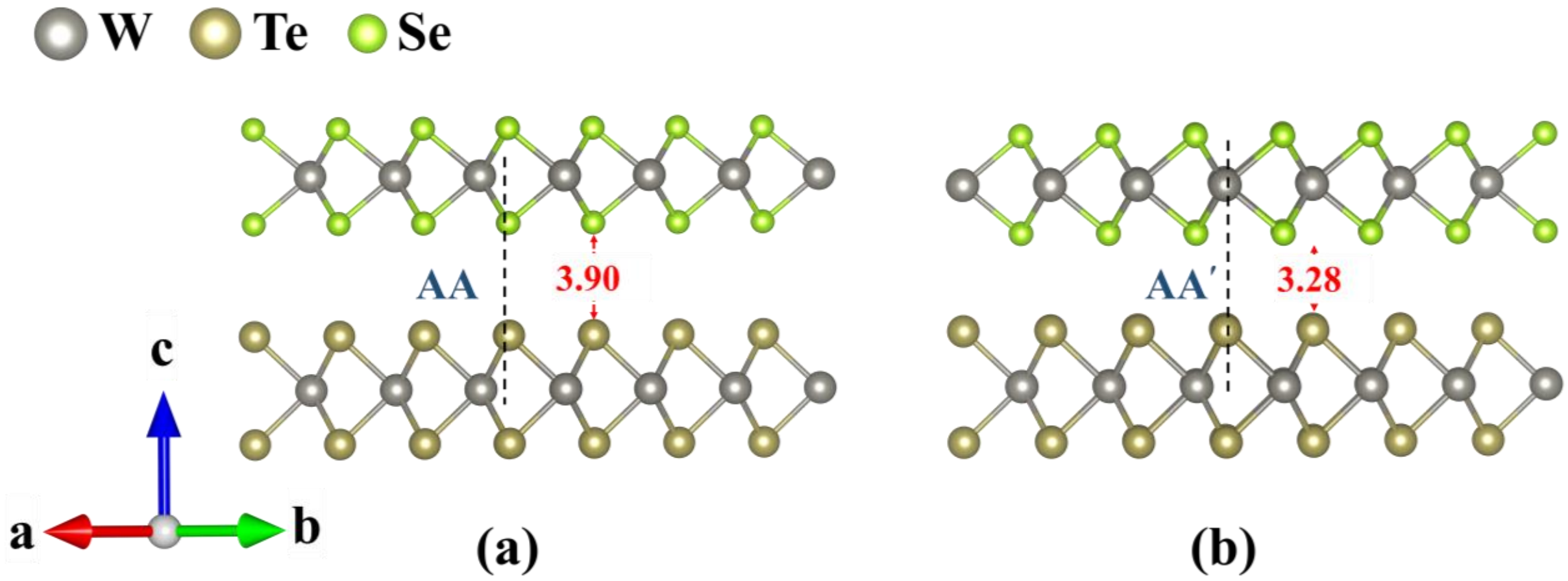

FIG 1. Stacking sequences of $WSe_2/WTe_2$ hetrobilayer (a) AA configuration, (b) AA′ configuration

## III. RESULTS

### A. Structural and electronic properties of $WSe_2$, $WTe_2$ monolayer and $WSe_2/WTe_2$ hetrobilayer

The optimized crystal structures of $WSe_2$ and $WTe_2$ monolayers are shown in Fig. S2. The optimized monolayers exhibit 2H phase with lattice parameters $a = b = 3.32$ Å for $WSe_2$ and $a = b = 3.56$ Å for $WTe_2$. Both structures exhibit in-plane lattice symmetry and out-of-plane mirror symmetry. The monolayer $WSe_2$ and $WTe_2$ have been synthesized experimentally and exhibit remarkable stability. The slight difference in lattice parameters results lattice mismatch of approximately 7.2%, which can introduce interfacial strain during the formation of $WSe_2/WTe_2$ heterobilayer. This lattice mismatch has been effectively accommodated in the heterobilayer without inducing significant structural distortion. To determine the most stable stacking configuration, two high symmetry stacking configurations, namely *AA* and *AA′*, has been considered to determine the energetically preferred stacking configuration. In the AA stacking, the chalcogen atoms of both layers are vertically aligned, whereas in the *AA′* configuration, the upper layer is laterally translated such that the transition metal atoms are positioned above the chalcogen atoms of the adjacent layer, as shown in Fig. 1(a) and (b). The optimized lattice parameter of *AA* and *AA′*-configuration has been found to be 3.400 Å and 3.402 Å, respectively. The calculated total energies indicate that the *AA′* stacking configuration is energetically more favorable than the *AA* stacking configuration. The relative stability of *AA′* configuration is further supported by the shorter interlayer distance in the *AA′* configuration (3.28 Å) compared to the *AA* configuration (3.90 Å). Therefore, *AA′* configuration has been considered for further study. The calculated electronic band structure and density of states (DOS) calculation of monolayer $WSe_2$, monolayer $WTe_2$ and hetrobilayer $WSe_2/WTe_2$ are presented in Fig. S3. The calculated band structures indicates that the monolayer $WSe_2$ and $WTe_2$, exhibit valence band maxima and conduction band minima at K point, The band edges (VBM, CBM) are located at (-0.81, 0.74) for monolayer $WSe_2$, and (-0.54, 0.53) for monolayer $WTe_2$, leading to direct band gap of 1.55 eV and 1.07 eV, respectively, Fig. S3(a) and S2(b). While the $WSe_2/WTe_2$ heterobilayer exhibits the significantly reduced band gap of 0.70 eV, having band edges at (-0.42, 0.28), Fig. S3(c). The presence of chemically inequivalent Se and Te layers in heterobilayer breaks the out-of-plane mirror symmetry, which gives rise to interface charge redistribution and suggest the formation of intrinsic interfacial dipole, as shown in Fig. 2(b). The layer projected band structure of $WSe_2/WTe_2$ hetrobilayer demonstrate that the Valence Band Maximum (VBM) originates form $WTe_2$ layer and Conduction Band Minimum (CBM) originates form $WSe_2$ layer, Fig. 2(a). This spatial separation of the band edges confirms the formation of Type-II band alignment in $WSe_2/WTe_2$ hetrobilayer. Such band alignment suggests that the electrons are predominantly localized in $WSe_2$ and holes in $WTe_2$, thereby reducing electron–hole recombination probability and facilitating interlayer charge separation. The Fig. S3(d-f) exhibit the projected density of states (PDOS) of $WSe_2$ monolayer, $WTe_2$ monolayer and WSe2/WTe2 hetrobilayer. Although the pristine $WSe_2/WTe_2$ heterobilayer exhibits a nonmagnetic type-II semiconducting ground state, its pronounced interfacial electronic reconstruction provides an ideal platform for introducing magnetism through selective 3d transition-metal substitution. Motivated by this observation, $W$ atoms in the $WSe_2$ layer have been systematically substituted with V, Cr, Mn, Fe, and Co to investigate how the heterointerface influences the emergence of magnetic ordering, spin-dependent electronic reconstruction, and magnetic anisotropy.

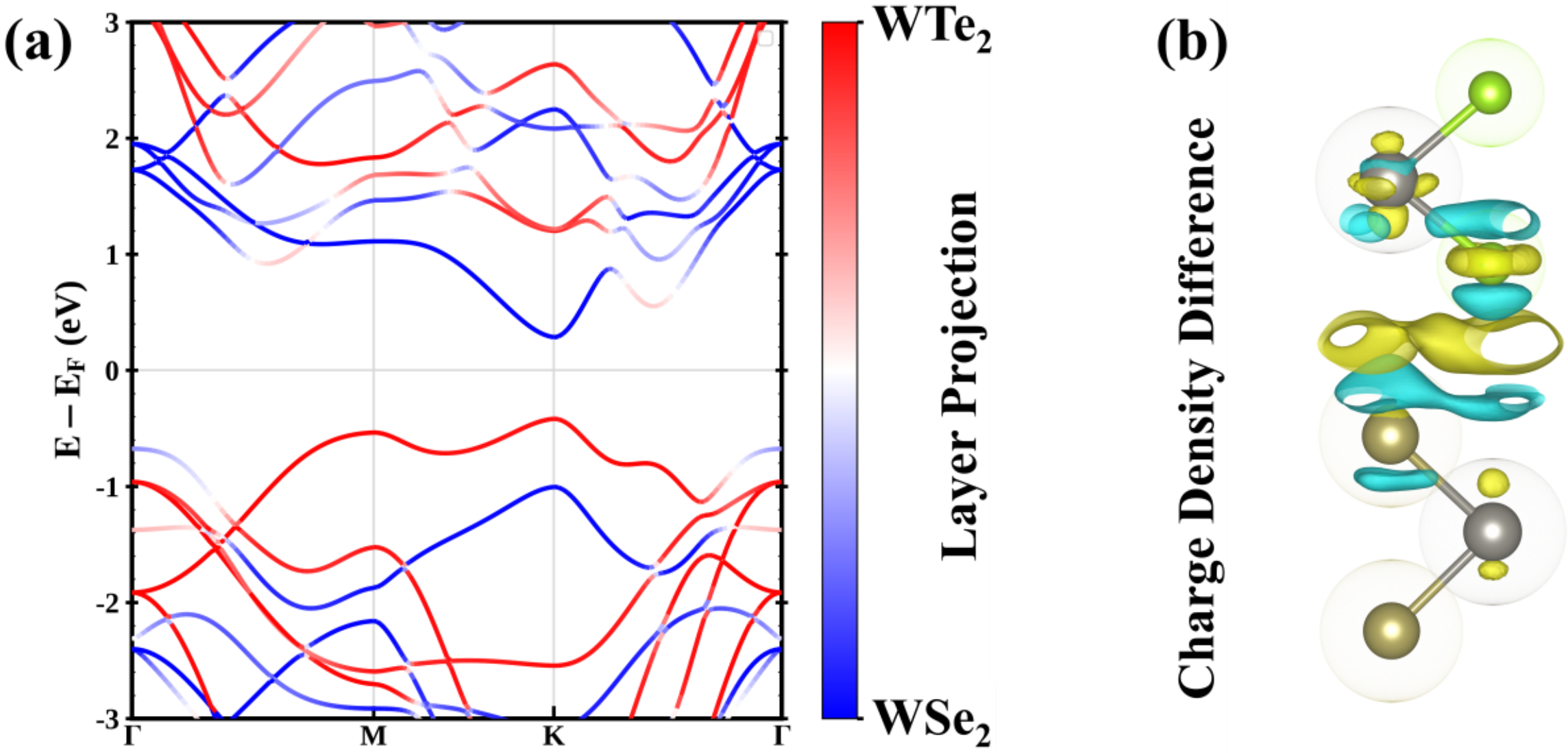

FIG 2. (a) Projected band structure and (b) Charge density difference isosurface of $WSe_2/WTe_2$ hetrobilayer (AA′ configuration)

### B. The $MSe_2/WTe_2$ hetrobilayer

The optimized $WSe_2/WTe_2$ heterobilayer exhibits the interlayer spacing ($h$), *W–Se* bond length, *W–Te* bond length, and corresponding bond angles $\angle_{Se-W-Se}$ and $\angle_{Te-W-Te}$ equals to 3.28 Å, 2.56 Å, 2.71 Å, 79.88°, and 87.31°, respectively. Substitution of W atoms with 3d transition metal atoms (V, Cr, Mn, Fe and Co) induces local structural reconstruction, as shown in Table SI. M- Se bond length ($d_{M\text{-}Se}$) varies from 2.05 Å (Co) to 2.63 Å (Cr), and interlayer distance varies from 2.86 Å (Co) to 3.11 Å (Cr). The variation in interlayer distance and bond lengths also influence the corresponding bond angles $\angle_{Se-M-Se}$. Substitution within one layer influence the structural parameter of adjacent $WTe_2$- layer, as confirmed by the notable change in W-Te bond length and bond angle, as shown in Table SI. The stability of these $MSe_2/WTe_2$ heterobilayer, has been assessed by the calculating formation energy per atom ($E_f$), as follows,

$$E_f = \frac{(E_{\mathrm{MSe2/WTe2}} - E_M - N_W E_W - N_{Se} E_{Se} - N_{Te} E_{Te})}{N_{total}} \quad (5)$$

.

Where $E_{MSe2/WTe2}$ is the total energy of the $MSe_2/WTe_2$ structure, $E_M$, $E_W$, $E_{Se}$ *and* $E_{Te}$ are the total energies of isolated 3d transition metal atom, W, Se and Te atom, respectively, while $N_W$, $N_{Se}$, $N_{Te}$ are the number of W, Se and Te atoms, respectively and $N_{total}$ is the number of total atoms in the structure. Table I summarizes the formation energy values for $MSe_2/WTe_2$ heterobilayers. The negative formation energies indicate all the structures are energetically stable and can be experimentally realized under favorable experimentally conditions. Among all the investigated heterobilayers, $MnSe_2/WTe_2$ exhibits the lowest formation energy (−3.83 eV/atom), indicating $MnSe_2/WTe_2$ structure is relatively stable and energetically more favorable than other structures. Fig. S4 exhibits the comparative summary including interlayer spacing, formation energy and magnetic moment of $MSe_2/WTe_2$ heterobilayers. The Se-Te interface in $MSe_2/WTe_2$ heterobilayer plays a crucial role to modulating the electronic and magnetic properties. To elucidate the microscopic electronic interaction at the heterointerface, charge-density difference (CDD) and planar-averaged electrostatic potential (PAEP) analyses have been performed, as depicted in Fig. 3. Fig. 3(b), (d), (f), (h) and (j), reveals that charge redistribution occurs upon hetrobilayer formation as compared to isolated $MSe_2$ and $WTe_2$ monolayers. For $CrSe_2/WTe_2$, $MnSe_2/WTe_2$ and $FeSe_2/WTe_2$, localized charge accumulation and depletion are confined to the Se–Te interface, indicating strong interfacial electronic polarization. The planar-average electrostatic potential profiles, which exhibits the significant potential difference across the interface are shown in Fig. 3(a), (c), (e), (g) and (i). The potential difference calculated across the Te-Se interface is 4.00 eV, 3.87 eV, 3.60 eV, 3.64 eV, and 3.83 eV for $VSe_2/WTe_2$, $CrSe_2/WTe_2$, $MnSe_2/WTe_2$, $FeSe_2/WTe_2$, and $CoSe_2/WTe_2$, respectively. Combining the electrostatic potential difference with the interlayer distance, as shown in Table S1, yields the intrinsic built-in electric fields of 1.37 eVÅ$^{-1}$, 1.24 eVÅ$^{-1}$, 1.19 eVÅ$^{-1}$, 1.20 eVÅ$^{-1}$ and 1.34 eVÅ$^{-1}$ for $VSe_2/WTe_2$, $CrSe_2/WTe_2$, $MnSe_2/WTe_2$, $FeSe_2/WTe_2$, and $CoSe_2/WTe_2$, respectively. The resulting intrinsic electric field is expected to reconstruct the electronic and magnetic behavior. The interface-induced electrostatic reconstruction establishes a nonmagnetic type-II semiconducting platform in pristine $WSe_2/WTe_2$. Introducing 3*d* transition-metal atoms into the $WSe_2$ layer perturbs the local crystal-field environment and strengthens the hybridization between the localized M 3*d* and Se 4*p* orbitals, resulting in exchange splitting and the emergence of spin-polarized electronic states. Consequently, the heterobilayers develop finite magnetic moments, as shown in Table I.

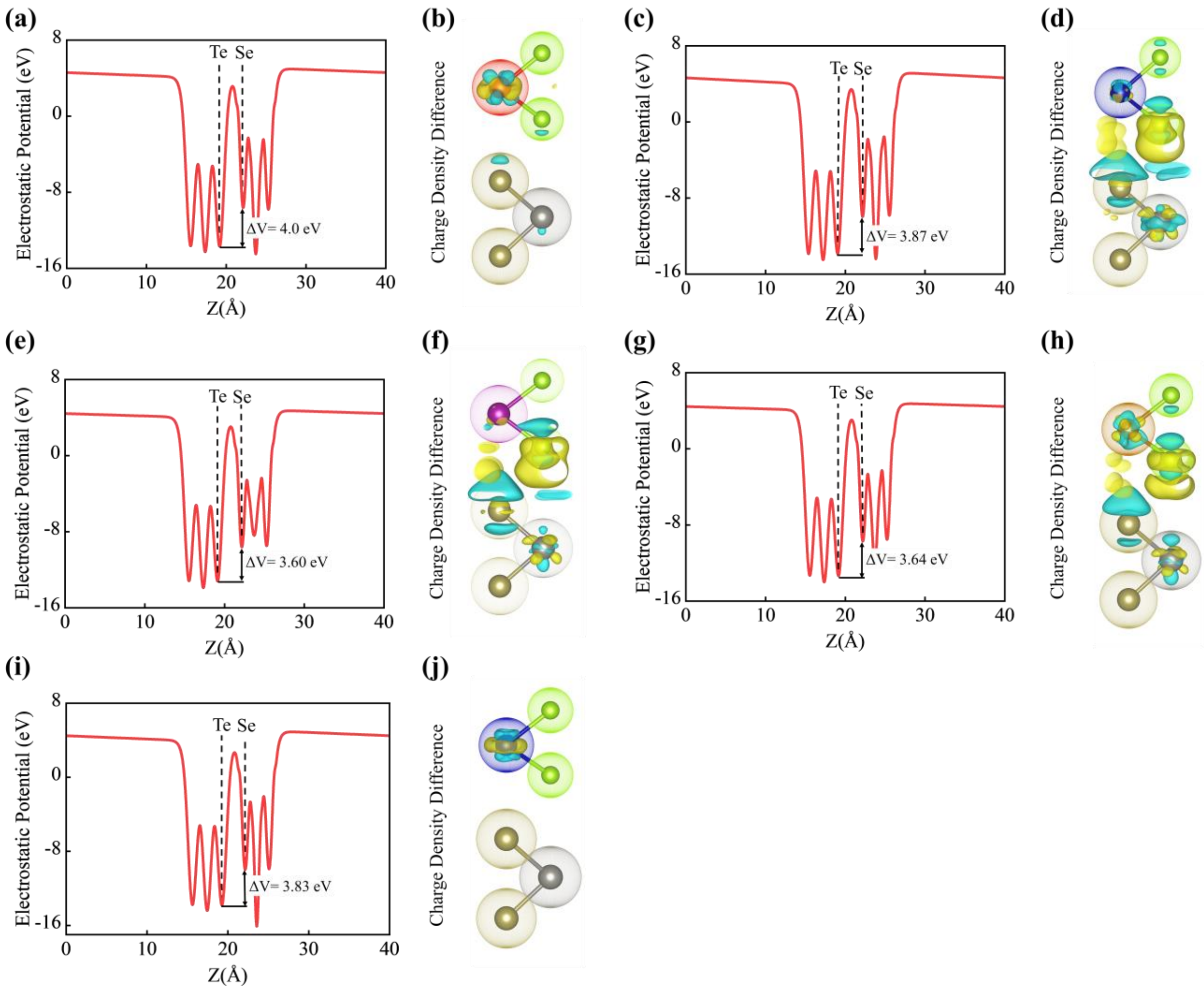

FIG 3. Planar average electrostatic potential profile and charge density difference of (a) and (b) $VSe_2/WTe_2$, (c) and (d) $CrSe_2/WTe_2$, (e) and (f) $MnSe_2/WTe_2$, (g) and (h) $FeSe_2/WTe_2$, and (i) and (j) $CoSe_2/WTe_2$ hetrobilayer

Table I. The Hubbard Value U, Formation energy per atom ($E_f$) and magnetic moment ($M$) of $MSe_2/WTe_2$ hetrobilayer.

| Structure | U (eV) | $E_f$ (eV) | Magnetic Moment M ($\mu_B$) |
|---|---|---|---|
| **$VSe_2/WTe_2$** | 5.08 | -3.82 | 1.40 |
| **$CrSe_2/WTe_2$** | 5.90 | -3.67 | 2.78 |
| **$MnSe_2/WTe_2$** | 4.84 | -3.83 | 3.00 |
| **$FeSe_2/WTe_2$** | 5.06 | -3.76 | 3.10 |
| **$CoSe_2/WTe_2$** | 7.77 | -3.58 | 0.58 |

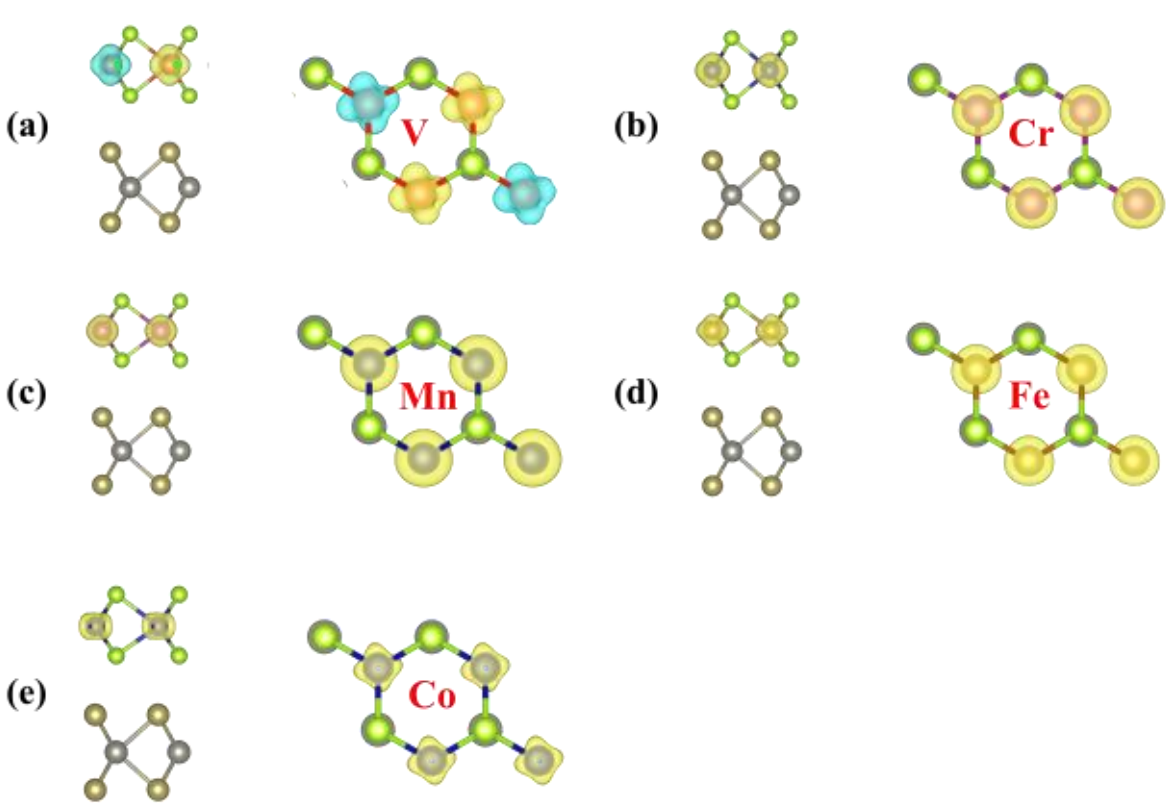

FIG 4. Spin density ($\rho_{up} - \rho_{down}$) isosurface for (a) $VSe_2/WTe_2$, (b) $CrSe_2/WTe_2$, (c) $MnSe_2/WTe_2$, (d) $FeSe_2/WTe_2$, and (e) $CoSe_2/WTe_2$ hetrobilayer structure having isosurface value 0.042.

A systematic comparison of the ferromagnetic (FM) and antiferromagnetic (AFM) configurations reveals that $VSe_2/WTe_2$ is antiferromagnetic where $CrMSe_2/WTe_2$, $MSe_2/WTe_2$, $FeSe_2/WTe_2$, and $CoSe_2/WTe_2$ exhibits ferromagnetic ground state. Fig. 4 exhibits the spin density difference isosurface ($\rho_{up} - \rho_{down}$) of $MSe_2/WTe_2$ heterobilayer. The yellow and blue isosurface represent the excess of spin-up and spin-down population, respectively. The localization of spin density around the 3*d* transition metal atom reveals that magnetic moment is mostly originated from partially occupied 3*d* orbitals of the transition metal atom. The nonuniform distribution of the spin density difference exhibits anisotropic characteristic, reflecting the directional dependence nature of magnetization. This anisotropy originates in the system due to crystal-field splitting and hybridization between the transition metal 3*d* and Se 4*p* orbitals, which modifies the electron occupancy of the $d_{xy}$, $d_{xz}$, $d_{yz}$, $d_{x^2-y^2}$ and $d_{z^2}$ states.

The microscopic origin of the magnetic behavior has been further investigated by spin-polarized projected density of states (PDOS), as shown in Fig. 5. The antiferromagnetic ground state of $VSe_2/WTe_2$ hetrobilayer, results in degenerates spin-up and spin-down electronic states, which indicates that the total magnetization of the system is zero, although the V atom carries a local magnetic moment of 1.40 $\mu_B$,

$$\mu_B \int [N_\uparrow(E) - N_\downarrow(E)]dE = 0 \quad (6)$$

Where, $N_\uparrow(E)$ and $N_\downarrow(E)$ are the spin-up and spin-down, density of states at energy E, respectively. In contrast, the ferromagnetic heterobilayers $CrSe_2/WTe_2$, $MnSe_2/WTe_2$, $FeSe_2/WTe_2$, and $CoSe_2/WTe_2$ displays significant spin asymmetry. The spin asymmetry in these system in primary originates from 3*d* states and its hybridization with surrounding Se 4*p* states, while the W 5*d* and Te 5*p* states remain nearly spin symmetric. The significant contribution of Se 4*p* orbitals near the Fermi level indicates that the low-energy electronic states are strongly governed by the local electronic environment of 3*d* transition metal atom. The electronic state near the Fermi level determines the electronic transport properties. The dominant contribution of Se 4*p* states near the Fermi level suggests that the low-energy electronic transport of these heterobilayers is largely dominated by the Se sublattice. The coexistence of transition-metal 3*d* states and Se 4*p* states near the Fermi level further indicates the strong *p-d* hybridization between the transition-metal 3d and Se 4*p* orbitals. Consistently, Fig. 4 shows that the spin density remains localized around the 3d transition metal atoms, and PDOS, Fig. 5, exhibits pronounced spin asymmetry in Se 4*p* states. This behavior suggests that the ligand polarization induced by electronic coupling between transition metal atom's 3*d* orbitals and surrounding Se 4*p* orbitals. Fig. 5(c) and 5(d), reveals that $MnSe_2/WTe_2$, $FeSe_2/WTe_2$ heterobilayer exhibit strong spin asymmetry near the Fermi level, where the spin-up states contribute mostly near Fermi level, while the - spin-down states develops a finite band gap. The absence of spin-down states at the Fermi level indicates that electrical conduction occurs exclusively through the spin-up channel, which confirms the spin-selective nature of $MnSe_2/WTe_2$, $FeSe_2/WTe_2$ heterobilayer. The spin polarized electronic band structures of $MSe_2/WTe_2$ hetrobilayer are presented in Fig. S5. The electronic band structure of $VSe_2/WTe_2$, $CrSe_2/WTe_2$, and $CoSe_2/WTe_2$ hetrobilayer cross the Fermi level for both spin-up and spin-down channels, which represent metallic behaviors of these systems, Fig. S5(a), (b) and (c). In contrast, $MnSe_2/WTe_2$ and $FeSe_2/WTe_2$ hetrobilayer exhibit spin dependence electronic behavior. In these systems, spin-up bands cross the Fermi level whereas the spin-down's valence band and conduction band exhibit the band gap of 0.94 eV and 0.37 eV, for $MnSe_2/WTe_2$ and $FeSe_2/WTe_2$ respectively, thereby establishing half-metallic ferromagnetic ground states.

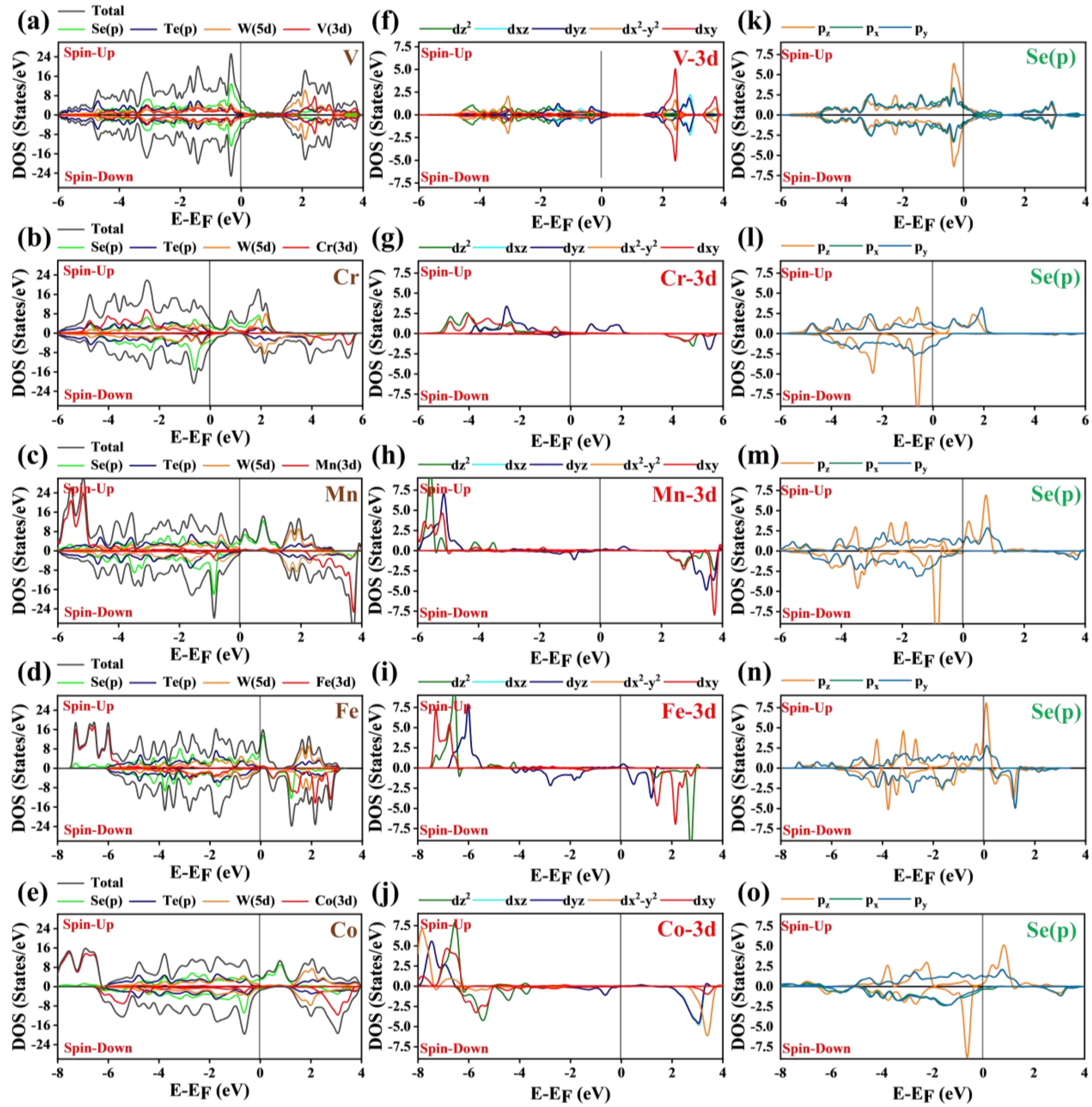

FIG 5. Partial Density of States (PDOS) of (a) $VSe_2/WTe_2$, (b) $CrSe_2/WTe_2$, (c) $MnSe_2/WTe_2$, (d) $FeSe_2/WTe_2$, and (e) $CoSe_2/WTe_2$ hetrobilayer 2×2×1 supercell with orbital-resolved PDOS of (f-j) 3d of transition metal atom and (k-o) 4p of Se.

To elucidate the microscopic origin of the spin dependent electronic states, the layer projected spin-resolved band structures of the half-metallic $MnSe_2/WTe_2$ and $FeSe_2/WTe_2$ heterobilayers are presented in Fig. 6. For $MnSe_2/WTe_2$ hetrobilayer, the metallic band crossing the Fermi level for spin-up state originate form the contribution of $MnSe_2$ layer, particularly from Se, Fig. 6(a). However, for spin-down, VBM and CBM has been originates from the contribution of $WTe_2$ layer, Fig. 6(b). These result shows that interfacial coupling induces layer dependent half-metallic electronic state. A similar electronic reconstruction is observed for $FeSe_2/WTe_2$ heterobilayer, where spin-up metallic state has been derived by the contribution of $FeSe_2$ layer at the Fermi level, Fig. 6(c). However, in the spin-down electronic state, the VBM has been formed by the $WTe_2$ layer, whereas the CBM has been formed by the $FeSe_2$ layer. This layer-selective localization of the band edges originates from the interface-induced electrostatic reconstruction, which shifts the relative band energies of the constituent layers without destroying the intrinsic half-metallic nature of the $FeSe_2$ layer. Consequently, the spin-up channel remains metallic, whereas the spin-down channel develops a type-II band alignment.

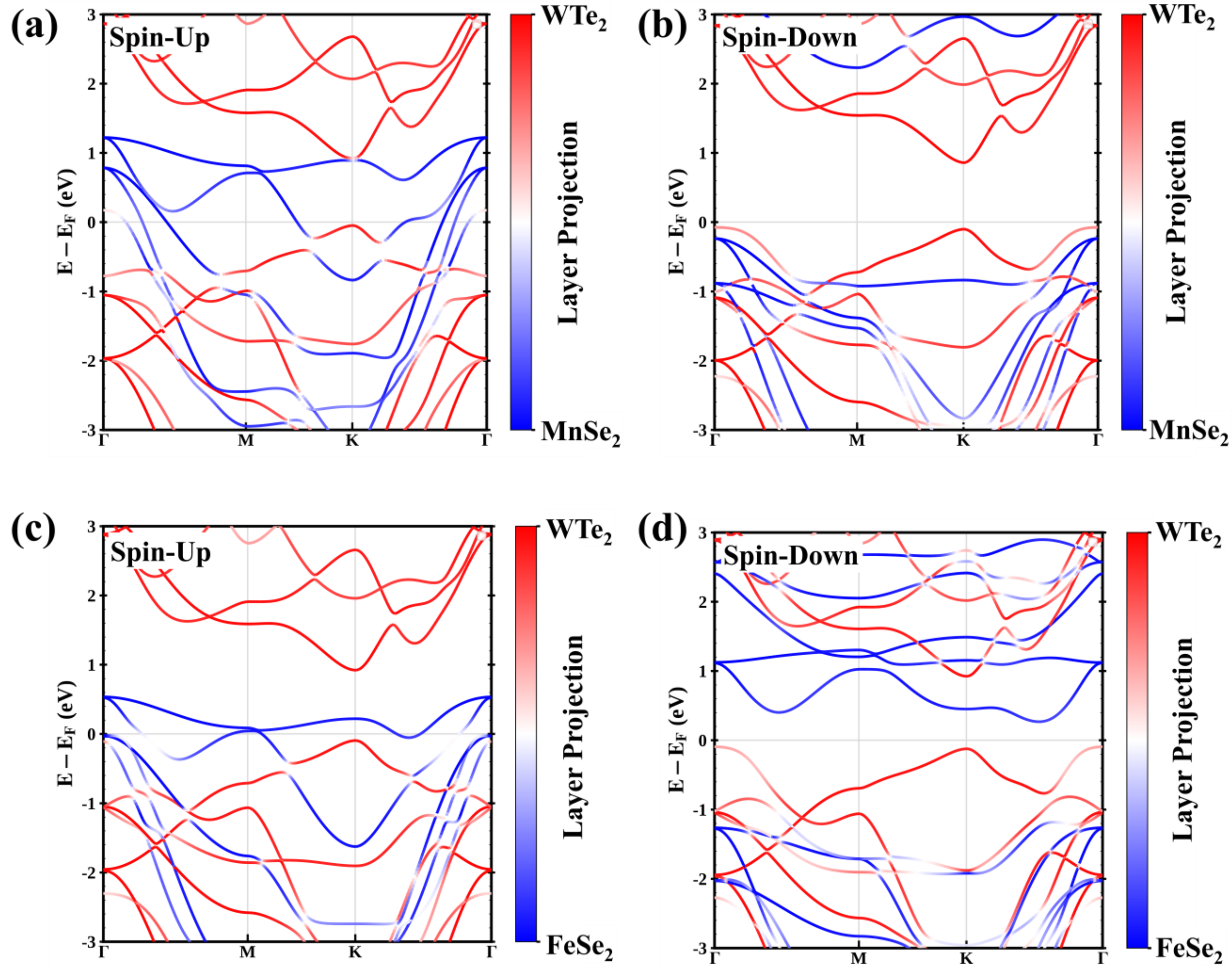

FIG 6. Spin-projected band diagram of (a) and (b) $MnSe_2/WTe_2$, and (c) and (d) $FeSe_2/WTe_2$

The degree of spin polarization has been further evaluated to quantify the spin-filtering capability of the $MSe_2/WTe_2$ heterobilayer, which has been calculated utilizing the following relation [29],

$$P = \frac{N_{\uparrow}(E_F) - N_{\downarrow}(E_F)}{N_{\uparrow}(E_F) + N_{\downarrow}(E_F)} \times 100\% \quad (7)$$

Where, $N_{\uparrow}(E_F)$ and $N_{\downarrow}(E_F)$ is the population of spin-up and spin-down density of states at the Fermi level, respectively. The calculated values of degree of spin polarization ($P$) for $MSe_2/WTe_2$ structures have been summarized in Table II. As $VSe_2/WTe_2$ exhibit antiferromagnetic nature, identical spin-up and spin-down density of states has been obtained at the Fermi level, which results in zero spin polarization. The $CrSe_2/WTe_2$ structure exhibits nearly equal contributions from spin-up and spin-down states at the Fermi level, with spin polarization of −1.41%, where the negative sign indicates that the spin down state is highly populated at Fermi level as compared to the spin up state. However, $MnSe_2/WTe_2$ and $FeSe_2/WTe_2$ exhibit complete absence of spin-down states at the Fermi level, which results in 100% spin polarization.

TABLE II.
Degree of spin polarization with spin up and spin down states at Fermi level for $MSe_2/WTe_2$ structure.

| Dopant | $N_{\uparrow}(E_F)$ | $N_{\downarrow}(E_F)$ | P (%) |
|---|---|---|---|
| V | 5.69 | 5.69 | 0 |
| Cr | 5.24 | 5.39 | -1.41 |
| Mn | 4.1 | 0 | 100 |
| Fe | 11.64 | 0 | 100 |
| Co | 3.811 | 1.53 | 42.70 |

### C. Magnetic Exchange Interaction

To quantify the magnetic coupling between neighboring transition-metal atoms, the magnetic exchange energy has been calculated by utilizing the following relation,

$$\Delta E = E_{AFM} - E_{FM} \quad (4)$$

Where, ($E_{AFM}$) is the total energy of the AFM state, and ($E_{FM}$) is the total energy of the FM state. The size of supercell has been doubled in x- and y-direction to evaluate the magnetic exchange interaction. The calculated magnetic exchange energy (*ΔE*), exchange coupling constant (*J*), magnetic coupling nature, nearest-neighbor TM–TM distance ($d_{TM-TM}$), TM–Se–TM bond angle $\angle_{\mathrm{M-Se-M}}$ has been summarized in Table III. The temperature dependent magnetic properties have been calculated utilizing Monte-Carlo simulation and Heisenberg model, utilizing exchange coupling constant obtained by equation (4). Table III displays the calculated exchange coupling constants. The $CrSe_2/WTe_2$, $MnSe_2/WTe_2$, $FeSe_2/WTe_2$ and $CoSe_2/WTe_2$ structures exhibit the positive exchange constant, which indicate that ferromagnetic phase is favorable in these structures, while the negative exchange coupling constant ($J$ = -18.46 meV), indicates anti-ferromagnetic phase is energetically more favorable for $VSe_2/WTe_2$ structure. Fig. 7 displays the temperature dependence of the normalized magnetic moment and specific heat for the $MSe_2/WTe_2$ hetrobilayer. The magnetization value decreases with increasing temperature due to thermal spin fluctuations until it reaches at the magnetic phase transition temperature, which indicates the loss of long-range magnetic order. The specific heat exhibits a strong peak at the magnetic phase transition temperature. The calculated value of magnetic phase transition temperature has been shown in Table III. Among all the structures, $CoSe_2/WTe_2$ hetrobilayer exhibits highest Curie temperature (273.87 K). Notably, $CoSe_2/WTe_2$ exhibits the highest Curie temperature 273.87 K, and exhibit the smallest magnetic moment among the structures. This behavior show that the magnitude of the exchange interaction is not directly governed by the local magnetic moment alone, but instead depends on the exchange pathway and the underlying electronic structure The structural geometry exhibit M–Se–M bonds, with bond angles ($\theta_{\text{M-Se-M}}$) ranging from 87.74° to 92.96°, as displayed in Table III, representing the edge-sharing transition-metal chalcogenide network. For these type of structures, the dominant magnetic exchange pathway can be described through M(3*d*)-Se(4*p*)-M(3*d*). According to the Goodenough-Kanamori-Anderson (GKA) rule, when the angle between magnetic atoms through a nonmagnetic atom is close to 90°, the exchange interaction primarily favors ferromagnetic coupling but remains highly sensitive to orbital occupancy and orbital overlap of magnetic atom and mediated nonmagnetic atom [30,31]. A relatively small variations in the electronic configuration of the transition-metal atoms can significantly modify the balance between competing ferromagnetic and antiferromagnetic exchange interactions. In $CrSe_2/WTe_2$, $MnSe_2/WTe_2$, and $CoSe_2/WTe_2$ hetrobilayer, the transition metal atom induces negatively spin polarization on the neighbor Se-atom and Se atom exhibits the antiparallel magnetic moment of -0.757 $\mu_B$, -0.727 $\mu_B$ and -0.676 $\mu_B$, respectively, which points out the possibility of ligand mediated exchange.

TABLE III.
Exchange Energy $\Delta E$, Magnetic coupling, distance between dopant $d_{TM-TM}$ and angle between dopant via Se ($\theta_{TM-Se-TM}$) for $MSe_2/WTe_2$ structures

| Dopant | $d_{\text{TM-TM}}$ (Å) | $\theta_{\text{M-Se-M}}$ (°) | ΔE (meV) | J (meV) | Coupling | $T_{C/N}$ (K) |
|---|---|---|---|---|---|---|
| V | 3.53 | 85.73 | -160.65 | -18.46 | AFM | 136.48 |
| Cr | 3.51 | 82.96 | 82.24 | 2.75 | FM | 72.91 |
| Mn | 3.59 | 87.02 | 84.47 | 2.35 | FM | 65.07 |
| Fe | 3.57 | 87.44 | 197.18 | 3.09 | FM | 79.65 |
| Co | 3.53 | 87.74 | 59.24 | 10.46 | FM | 273.87 |

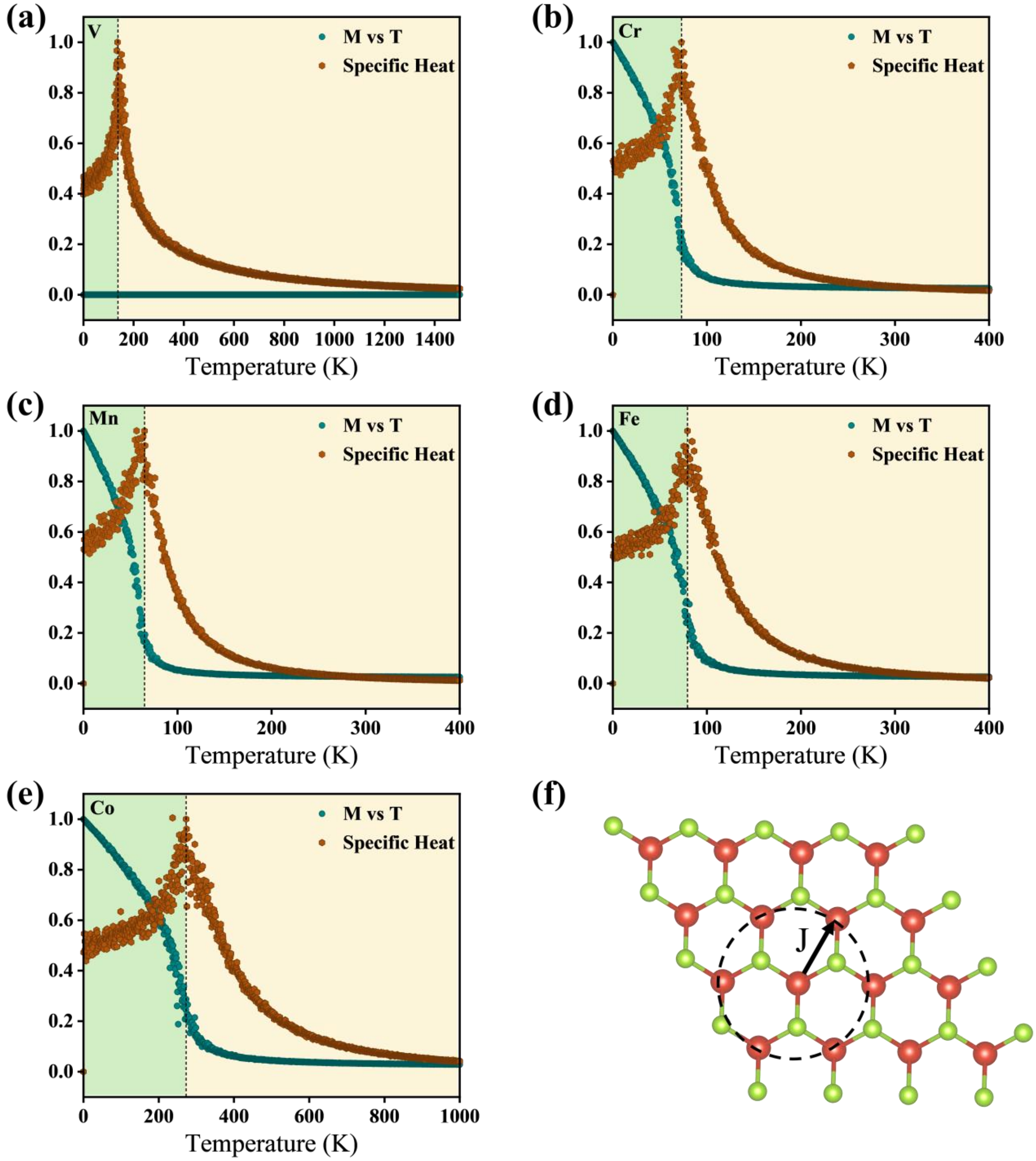

FIG 7. Magnetization and Specific heat in normalized unit as function of temperature for (a) $VSe_2/WTe_2$, (b) $CrSe_2/WTe_2$, (c) $MnSe_2/WTe_2$, (d) $FeSe_2/WTe_2$, and (e) $CoSe_2/WTe_2$ hetrobilayer system, (f) Exchange interaction range with nearest neighbors' atoms.

The PDOS indicates the electronic states near the Fermi level have significant Se 4*p* contribution, the substantial spin polarization on the Se atoms demonstrates that the ligand orbitals actively participate in the magnetic interaction rather than acting as passive structural bridges. However, $FeSe_2/WTe_2$ exhibits a comparatively weaker positive Se polarization (+0.179 $\mu_B$) and reduced exchange splitting, which indicates the existence of itinerant exchange interaction. The coexistence of large localized magnetic moments of 3*d* atoms together with spin polarization ligand (Se) indicates that the dominant exchange interaction remains localized and ligand mediated. The metallic states may introduce a secondary itinerant contribution to the magnetic response, particularly in the $VSe_2/WTe_2$, $FeSe_2/WTe_2$ and $CoSe_2/WTe_2$ heterobilayers where transition-metal 3d states exhibit finite occupancy near the Fermi level. The $MnSe_2/WTe_2$ heterobilayer exhibits the largest exchange splitting and a highly localized high-spin configuration, representing the most pronounced realization of localized ferromagnetic superexchange, whereas the Fe-based heterobilayer exhibits an intermediate localized–itinerant mechanism due to its partially occupied minority-spin 3*d* states near the Fermi level.

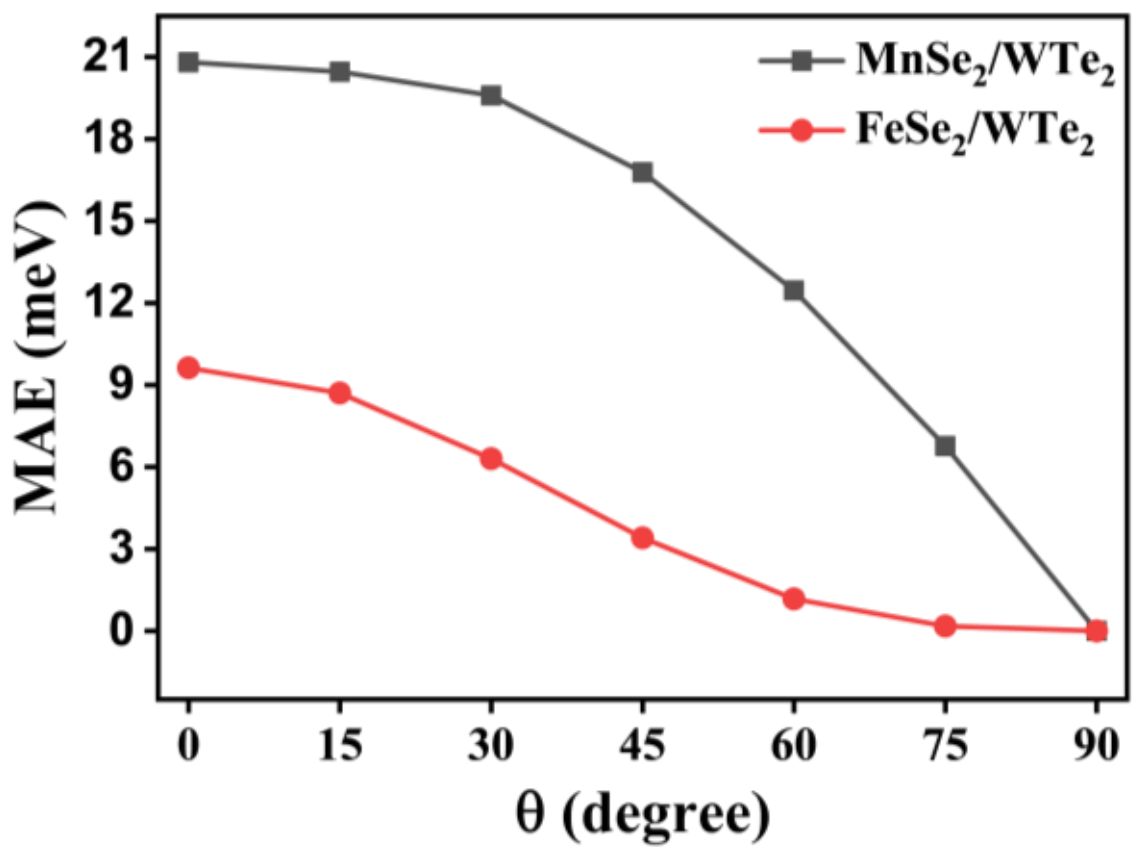

**FIG 8.** Magnetic Anisotropy Energy for $MnSe_2/WTe_2$ and $FeSe_2/WTe_2$ heterobilayers

### D. Magnetic Anisotropy Energy (MAE)

Magnetic anisotropy plays critical role in stabilizing long-range magnetic order in two-dimensional systems by suppressing the thermally excited spin fluctuations and breaking the continuous spin-rotational SU(2) symmetry, predicted by the Mermin-Wagner theorem. To determine the preferred magnetization direction, SOC enable calculation have been performed by rotating the magnetization direction from 0° (out-of-plane) to 90° (in-plane), in steps of 15°. Fig. 8, exhibits the angular dependence MAE for $MnSe_2/WTe_2$ and $FeSe_2/WTe_2$ hetrobilayer. The continuous reduction in MAE value, as the magnetization rotates from out of plane to in-plane direction, indicates that system preferred out-of-plane magnetization. The MAE value calculated utilizing the relation,

$$MAE = E_{parallel} - E_{perpendicular} \quad (8)$$

Where, $E_{parallel}$ and $E_{perpendicular}$ denotes the total energies for in-plane (90°) and out-of-plane (0°) configuration, respectively. The isolated $MnSe_2$ monolayer exhibits MAE of -1.12 meV, after forming $MnSe_2/WTe_2$ hetrobilayer, the MAE increases to 20.8 meV with magnetic easy axis transition from in-plane to out-plane. The $FeSe_2$ monolayer intrinsically exhibits out-plane easy axis with a MAE of 3.74 meV, which is further enhanced in $FeSe_2/WTe_2$ hetrobilayer to 9.63 meV. The substantial perpendicular magnetic anisotropy enhances the stability of the magnetic state against thermal spin fluctuations, whereas the half-metallic electronic structure of $MnSe_2/WTe_2$ and $FeSe_2/WTe_2$ hetrobilayer provides fully spin-polarized charge carriers. The formation of the $WTe_2$ heterointerface modifies the local electronic state through interfacial electronic reconstruction. As shown by the charge-density redistribution and electrostatic potential profile, Fig. 3. Such electronic reconstruction alters the anisotropic spin–orbit coupling between occupied and unoccupied 3d states, thus significantly enhances the MAE.

From a spintronic perspective, the enhanced perpendicular magnetic anisotropy with out of plane easy-axis exhibits significant potential. For example, the $FeSe_2/WTe_2$ hetrobilayer, simultaneously exhibits robust perpendicular magnetic anisotropy, half-metallicity and spin-resolved type- II band alignment making it a promising candite for high-density nonvolatile magnetic memory, spin-transfer-torque devices, and spin-filtering applications, where strong perpendicular magnetic anisotropy is needed for maintaining thermal stability while minimizing switching energy.

## IV. DISCUSSION

We have shown that heterostructure formation with the heavy TMD $WTe_2$, modifies the electronic and magnetic characteristics of $MSe_2$ (M = V, Cr, Mn, Fe and Co). The interfacial charge redistribution creates an intrinsic build in electric field at interface, which ranges from 1.19-1.37 $eVÅ^{-1}$, Fig. 3. This build-in-electric field reconstructs the ligand Se 4*p* orbital levels and modify the crystal field environment of the 3*d* transition metals (TM). Similar interface-induced electronic reconstruction has recently been reported [32], where Huang et al. have exhibited that a built-in electric field from adsorbed ions substantially increases ferromagnetic coupling and MAE in a 2D magnet. Zhu et al. have experimentally demonstrated that placing a 2D magnet ($CrPS_4$) on graphene can increase its Curie temperature to room temperature via interface charge transfer [33]. Rajawat et al. demonstrated that interface engineering in $\beta$-W/Py heterostructures substantially enhances spin pumping through strong spin–orbit coupling, underscoring the importance of interface effects in spintronics [34]. Our computed charge-density difference confirms significant electron depletion in $WTe_2$ and accumulation in $MSe_2$ at the interface, consistent with the formation of interface dipole, Fig. 3. This interfacial charge redistribution influences the local symmetry and crystal field at the TM sites. The reconstructed crystal-field splitting is further supported by the PDOS, TM 3$d$ states near the Fermi level hybridize more strongly with Se 4$p$ states, Fig. 5(a-e). Notably, small induced magnetic moments appear on Se equals to –0.757 $\mu_B$, –0.727 $\mu_B$, –0.676 $\mu_B$ for $CrSe_2/WTe_2$, $MnSe_2/WTe_2$, $CoSe_2/WTe_2$, respectively, which indicates ligand participation in exchange interaction. The fully spin-polarized, half-metallic states is directly visible in spin-polarized band structure. Such 100% spin polarization is particularly attractive for spin injection and filtering. In $FeSe_2/WTe_2$, spin-up channel remains metallic while spin-down channel reconstructs, the VBM originating from $WTe_2$ and CBM originating from $FeSe_2$, which exhibits half-metallic behavior with spin-resolved type-II band alignment, Fig. 6(d). Unlike conventional type-II semiconductor heterojunctions, where charge separation occurs identically for both spin channels, the present heterobilayers exhibit a spin-selective type-II alignment in which only the spin-down channel undergoes spatial separation of the band edges while the spin-up channel preserves metallic transport. This coexistence of half-metallicity and spin-resolved type-II alignment provides simultaneous spin filtering and spatial separation of spin-polarized carriers, a combination rarely reported in magnetic van der Waals heterostructures. This spin-resolved alignment suggests potential device concepts, in which the heterobilayer can act as a spin-selective diode or photodetector in which optical excitation or applied bias generates a spin-polarized current. For example, Dang et al. demonstrated a 2D spin-LED using a $CrI_3$/hBN/$WSe_2$ heterostructure, achieving circularly polarized electroluminescence of ~40% [35]. The microscopic mechanism responsible for magnetic ordering can be interpreted as the coexistence of itinerant and localized ligand-mediated exchange mechanism. Several metallic two-dimensional materials, such as $Fe_3GeTe_2$, $MnBiTe_4$, etc., the magnetic ordering can stabilize through combination of Ruderman-Kittel-Kasuya-Yosida (RKKY) interaction and superexchange interaction. The metallic electronic character of our system points toward existence of itinerant mechanism. However, the electronic and magnetic analyses consistently indicate to a ligand-mediated superexchange. Sasioglu et al. have demonstrated the coexistence of itinerant and localized mechanism, while exploring spin-gapped metallic half-Heusler compounds [36]. Furthermore, the heterobilayers exhibits significantly enhanced MAE for $MnSe_2/WTe_2$ (-1.12 meV to +20.8 meV) and $FeSe_2/WTe_2$ (+3.74 meV to 9.63 meV). This enhancement in MAE can be understood by second order spin–orbit perturbation theory [37]. The MAE can be expressed by utilizing the following relation:

$$\mathrm{MAE} \propto \xi^2 \sum_{o,u} \frac{|\langle\psi_u|\mathbf{L}_z|\psi_o\rangle|^2 - |\langle\psi_u|\mathbf{L}_x|\psi_o\rangle|^2}{E_u - E_o} \qquad (9)$$

Where, $\xi$ is the spin-orbit coupling constant, $\boldsymbol{L}$ represents the orbital momentum operator, and $\psi_o$ and $\psi_u$ represent the occupied and unoccupied electronic states, $E_u$ and $E_o$ represent the energy level of $\psi_u$ and $\psi_o$ states. The interface electric field modifies the energy of ligand (Se) and reorders the occupied and unoccupied 3$d$ levels by crystal field modification. Consequently, the interface-induced crystal-field reconstruction modifies the energy separation between occupied and unoccupied states. Tang et al. also observed similar behavior for $Fe_3GeTe_2$ that interfacing $Fe_3GeTe_2$ with $VI_3$ layer, significantly enhances the magnetic anisotropy [38]. The strong out-of-plane MAE in $MnSe_2/WTe_2$ suggests the magnetization can be switched with current-induced torques. $WTe_2$ itself has large spin Hall and Rashba–Edelstein effects, so an in-plane current could exert a spin–orbit torque on the interfacial $MnSe_2$ moments. Overall, the present results demonstrate that interface engineering provides a unified route for simultaneously tailoring magnetic exchange, magnetic anisotropy, and spin-dependent electronic reconstruction in van der Waals heterostructures. This unified mechanism explains the simultaneous emergence of robust ferromagnetism, enhanced perpendicular magnetic anisotropy, half-metallicity, and spin-resolved type-II band alignment, establishing heterobilayers as a promising model platform for multifunctional two-dimensional spintronic devices.

## V. CONCLUSIONS

In this study, a detailed first-principles investigation has been conducted to explore the structural, electronic and magnetic properties of $MSe_2/WTe_2$ (M = V, Cr, Mn, Fe, and Co) vdW hetrobilayer. The pristine $WSe_2/WTe_2$ heterobilayer AA′-configuration is energetically more favorable and exhibits type-II band structure, while transition-metal substitution induces diverse magnetic ground states ranging from antiferromagnetism in $VSe_2/WTe_2$ to robust ferromagnetism in Cr-, Mn-, Fe-, and Co-based heterobilayers. Charge-density-difference and electrostatic-potential analyses reveal pronounced interfacial charge redistribution accompanied by a strong intrinsic built-in electric field, which reconstructs the local crystal-field environment and enhances transition-metal 3d and Se(4p) hybridization. $MnSe_2/WTe_2$ displays half-metallic behavior with minority-spin band gaps of 0.94 eV, whereas $FeSe_2/WTe_2$ simultaneously exhibit half-metallicity and spin-resolved type-II band alignment, enabling fully spin-polarized transport together with spin-selective carrier separation. Furthermore, heterostructure formation significantly enhances the perpendicular magnetic anisotropy, including a spin-reorientation transition in $MnSe_2$ from an in-plane to an out-of-plane easy axis. The predicted half-metallicity, strong perpendicular magnetic anisotropy, and thermally stable magnetic phases, establish $MSe_2/WTe_2$ heterobilayers as promising candidates for next-generation two-dimensional spintronic device application.

## ACKNOWLEDGMENTS

Paras Poswal sincerely thank Dr. Ravindra Pandey for insightful suggestions, and continuous encouragement. Paras Poswal acknowledge NIT Patna and Ministry of Education (M.O.E) for fellowship. We acknowledge Department of Science and Technology (DST) and Ministry of Electronics and IT (MeitY) for providing access to High-Performance Computing (HPC) facility (PARAM Smriti, NABI Mohali) through National Supercomputing Mission (NSM).

# Supporting Information

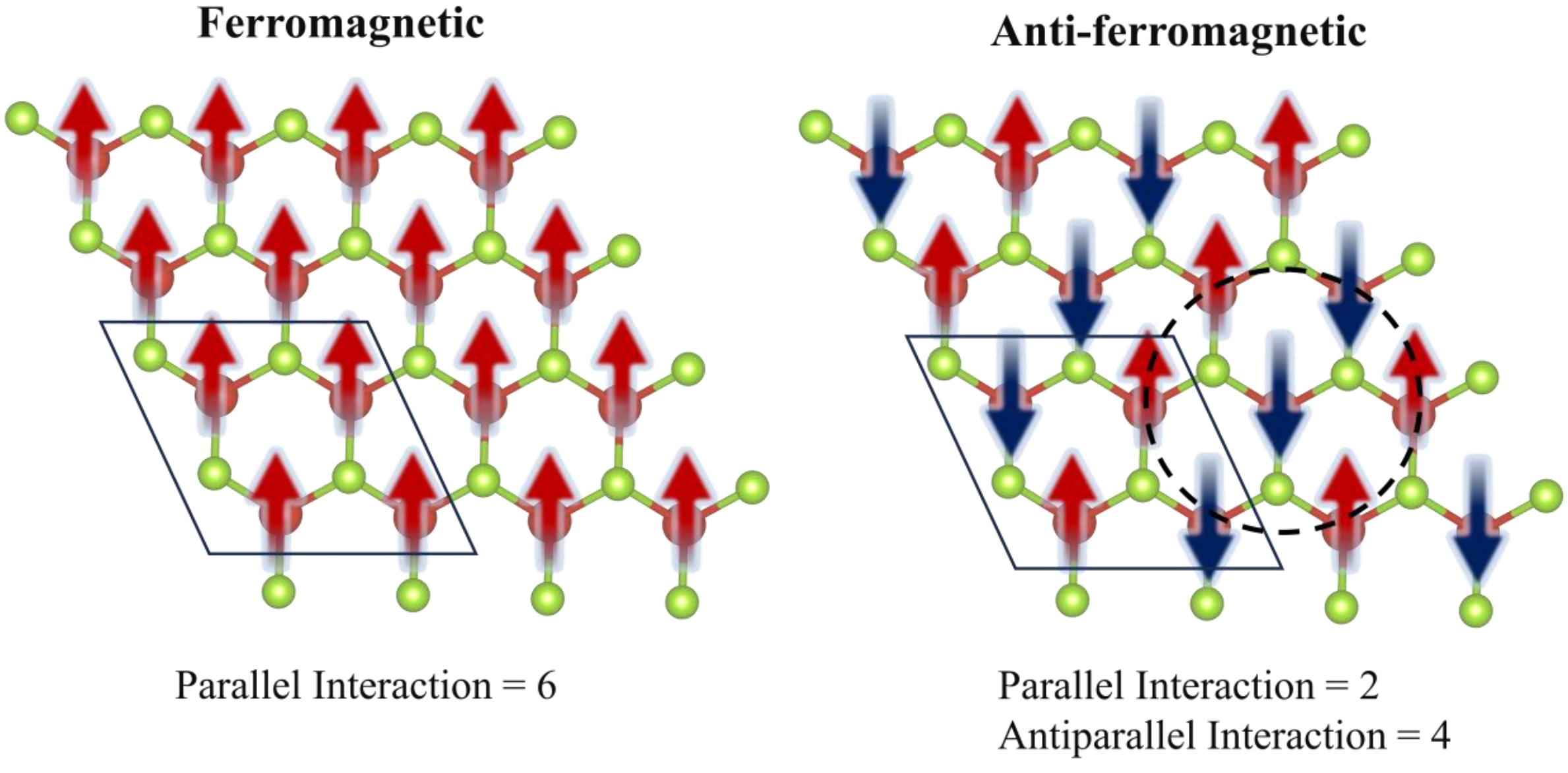

FIG S1. Spin arrangement in ferromagnetic and anti-ferromagnetic configuration, rectangular box represent the unit-cell used in total energy calculation.

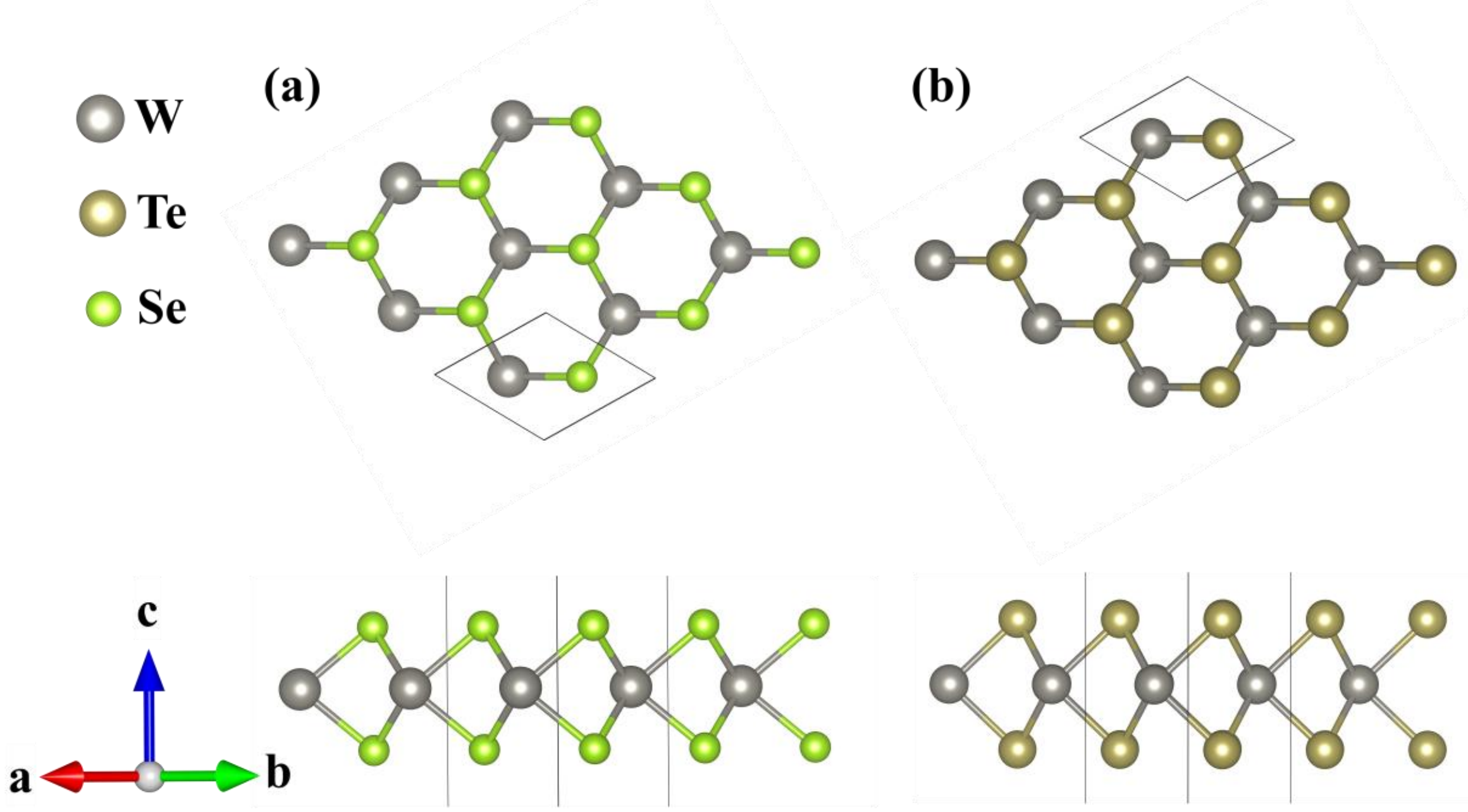

FIG S2. (a) Top view and side view of $WSe_2$ monolayer and (b) Top view and side view of $WTe_2$ monolayer.

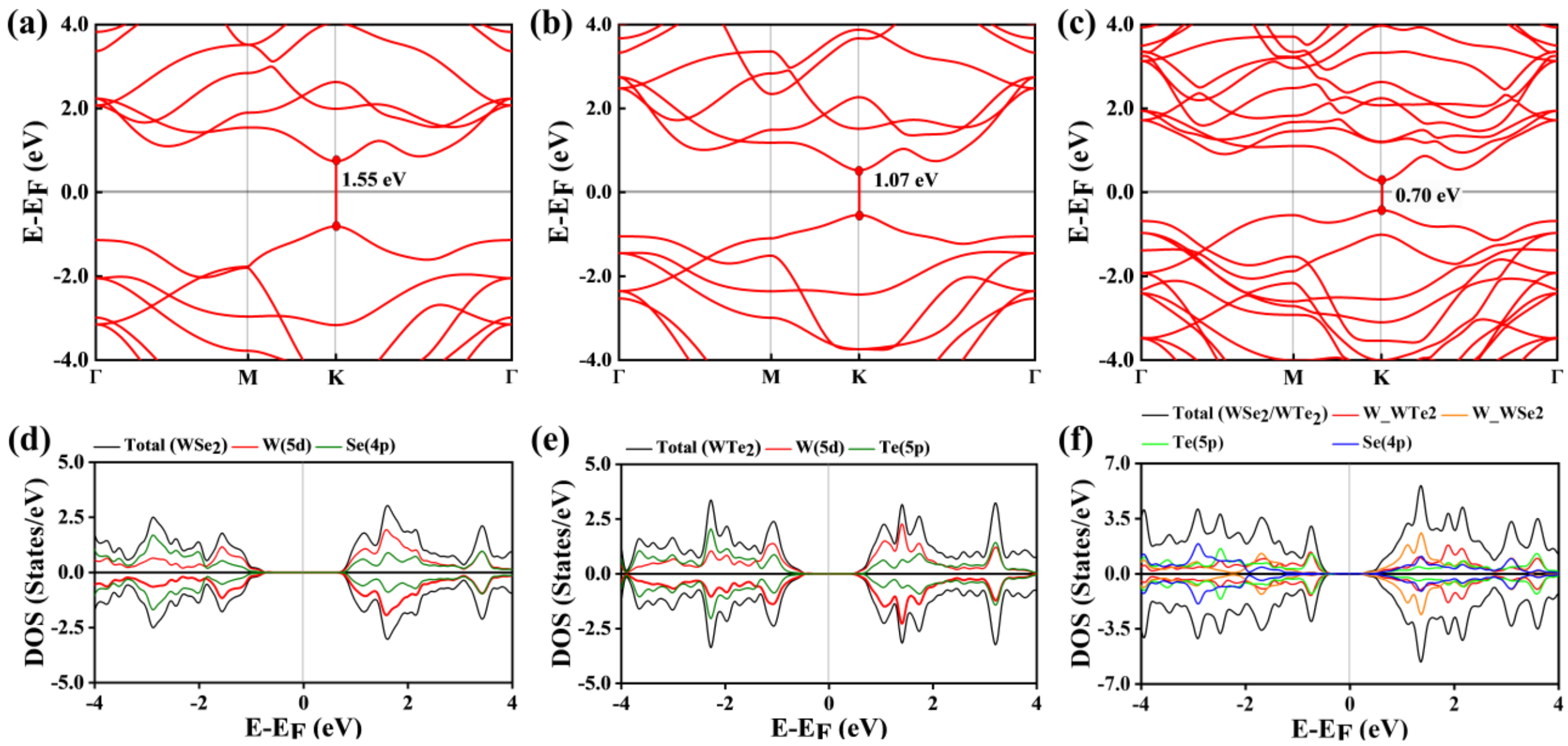

FIG S3. Band structure and DOS of (a) $WSe_2$ monolayer, (b) $WTe_2$ monolayer (c) $WSe_2$/ $WTe_2$ hetrobilayer (AA′ configuration)

Table SI. Structural parameters including interlayer spacing (h), bond length (d) and bond angle (θ) of $MSe_2/WSe_2$ hetrobilayer

| Structure | h(Å) | $d_{M\text{-}Se}$ (Å) | $d_{W\text{-}Te}$ (Å) | $\theta_{Se\text{-}M\text{-}Se}$ (°) | $\theta_{Te\text{-}W\text{-}Te}$ (°) |
|---|---|---|---|---|---|
| **$WSe_2/WTe_2$** | 3.28 | 2.56 | 2.71 | 79.88 | 87.31 |
| **$VSe_2/WTe_2$** | 2.93 | 2.58 | 2.73 | 76.35 | 83.57 |
| **$CrSe_2/WTe_2$** | 3.11 | 2.63 | 2.73 | 79.68 | 84.05 |
| **$MnSe_2/WTe_2$** | 3.02 | 2.59 | 2.74 | 74.40 | 81.90 |
| **$FeSe_2/WTe_2$** | 3.03 | 2.55 | 2.73 | 73.37 | 82.48 |
| **$CoSe_2/WTe_2$** | 2.86 | 2.05 | 2.73 | 72.26 | 83.61 |

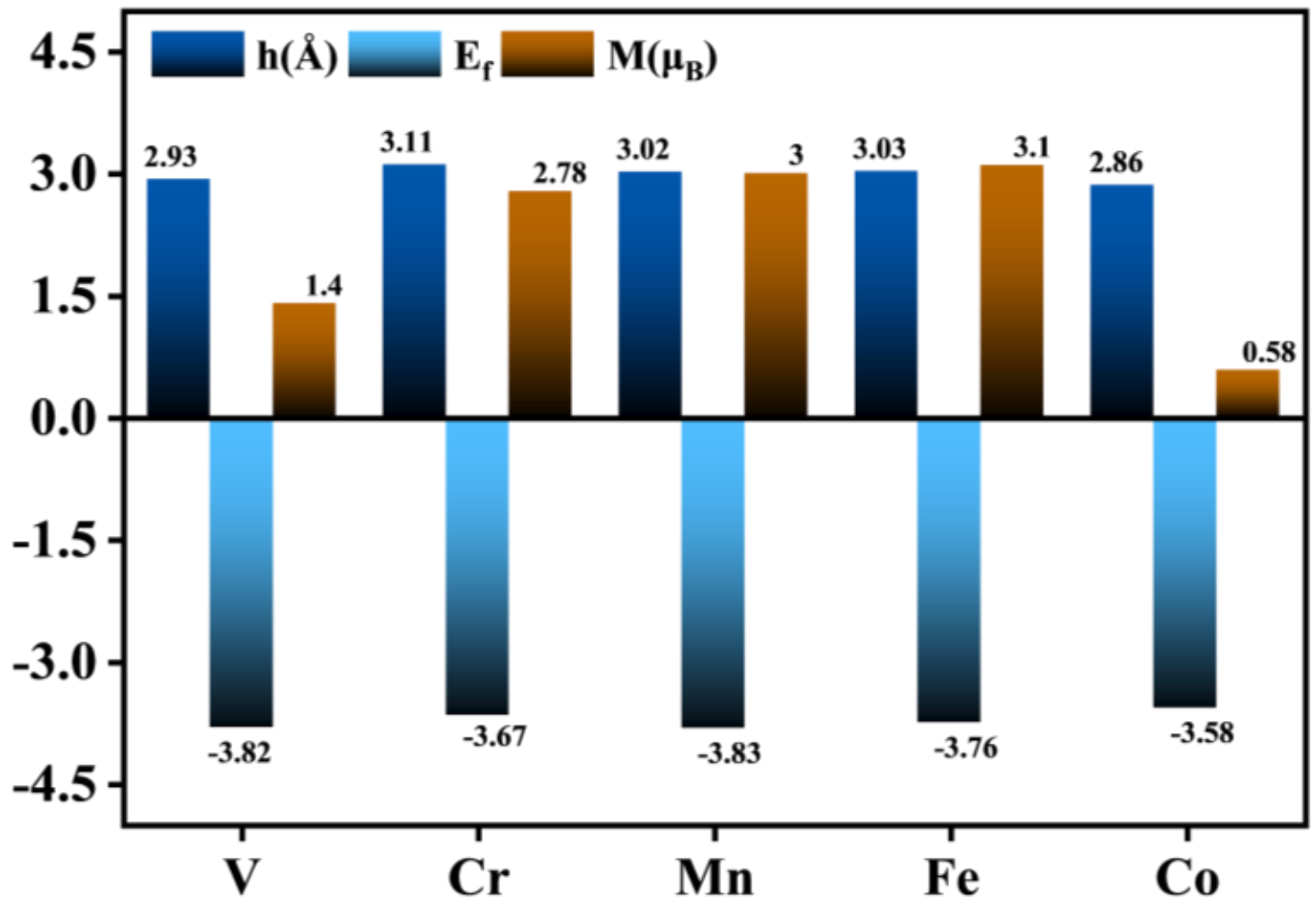

FIG S4. The graphical representation of interlayer spacing (h), Magnetic moment and Formation energy ($E_f$) of $MSe_2/WTe_2$ heterobilayer

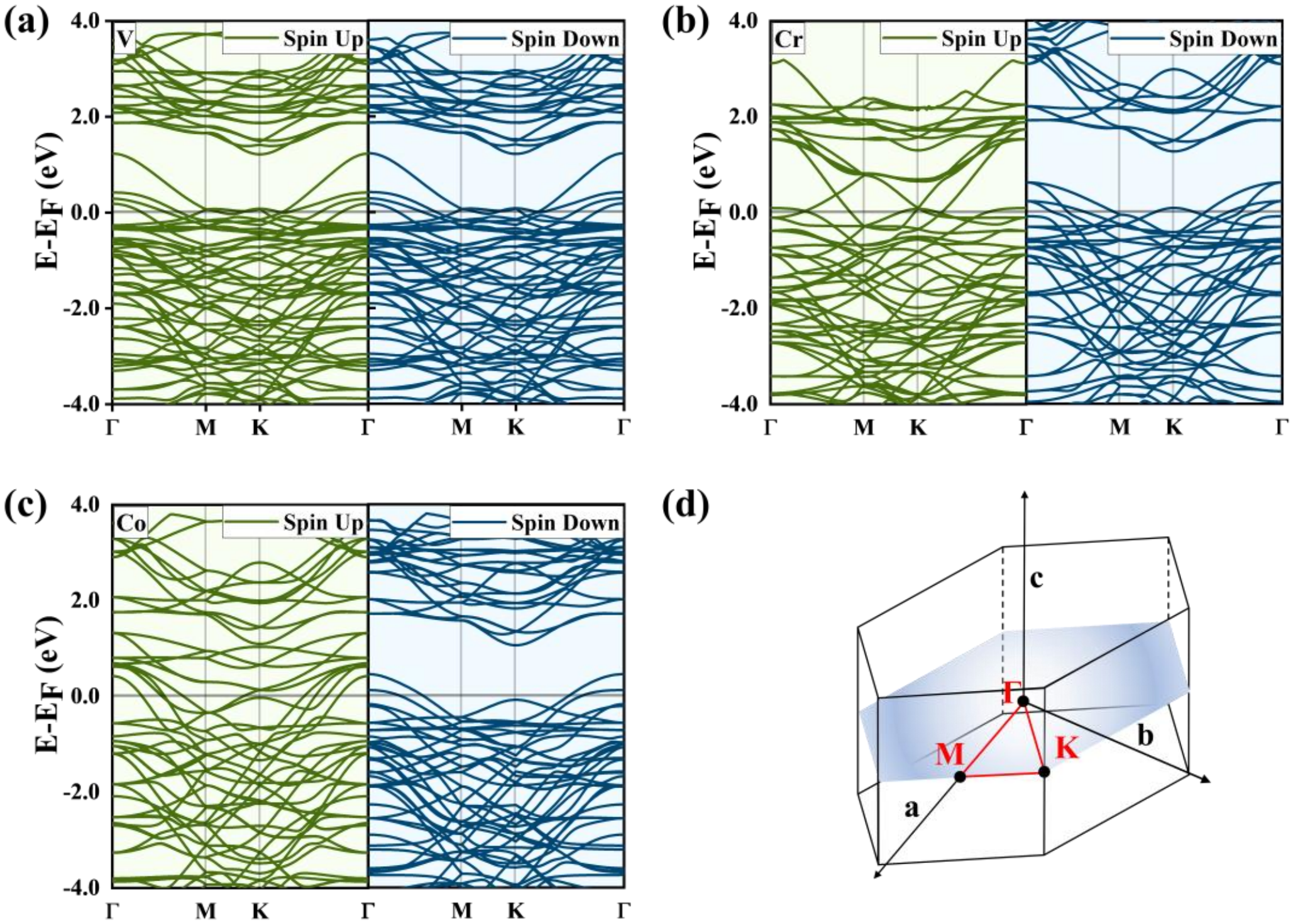

FIG S5. The band diagram of (a) $VSe_2/WTe_2$, (b) $CrSe_2/WTe_2$, (c) $CoSe_2/WTe_2$ hetrobilayer structure and (d) Brillouin zone with high- symmetry path